\documentstyle[10pt]{article}
\newcommand{\be}{\begin{eqnarray}}
\newcommand{\ee}{\end{eqnarray}}
\newcommand{\ba}{\begin{array}}
\newcommand{\ea}{\end{array}}
\newcommand{\tr}{\mbox{\rm tr}}

\begin{document}
%
%
\rightline{RUB-TPII-02/99}
\vspace{0.5cm}
\begin{center}

{\Large
Hard exclusive electroproduction of two pions and their resonances}
\footnote{Talk given at conference ``Particle and Nuclear Physics
with CEBAF at Jefferson Lab", Dubrovnik, 3-10 November 1998}
\\
\vspace{0.5cm}
{\bf  M.V.\ Polyakov}

\vspace{0.2cm}
{Petersburg Nuclear Physics
Institute, Gatchina, St.\ Petersburg 188350, Russia}\\
and\\
Institut f\"ur Theoretische Physik II,
Ruhr--Universit\"at Bochum, D--44780 Bochum, Germany

%
%
\end{center}
\vspace{0.5cm}

\begin{abstract}
We study the hard exclusive production of two pions in the virtual
photon fragmentation region with various invariant masses including the
resonance region. The amplitude is expressed in terms of two-pion light
cone distribution amplitudes ($2\pi$DA's). We derive dispersion
relations for these amplitudes, which enables us to fix them completely
in terms of $\pi\pi$ scattering phases and a few low-energy subtraction
constants determind by the effective chiral lagrangian. Quantitative
estimates of the resonance as well $\pi\pi$ background DA's at low
normalization point are made. We also prove certain new soft pion
theorem relating two-pion DA's to the one-pion DA's.
Crossing relations between $2\pi$DA's and parton distributions
in a pion are discussed.

\noindent
We demontrate that by studying the \underline{shape} of the $\pi\pi$
mass spectra (not absolute cross section!) in a diffractive
electroproduction one can extract the deviation of the meson
($\pi,\rho,$~etc.) wave functions from their asymptotic form
$6 z(1-z)$ and hence to get important information about the structure
of mesons. We suggest (alternative to S\"oding's) parametrization
of $\pi\pi$ spectra which is suitable for large photon virtuality.

\end{abstract}
\vspace{0.2cm}
%
%
%

\noindent
The aim of this talk is to outline the approach to two-pion
hard electroproduction developed in \cite{MVP98}. Here we summarize
the main results, details can be found in \cite{MVP98}.

Hard exclusive electroproduction of mesons is a new kind of
hard processes calculable in QCD.
In \cite{CFS,Rad} was shown that a number of exclusive
processes of the type:

\be
\gamma_L^\ast(q)+ T(p) \to F(q') + T'(p')\;,
\label{proc1}
\ee
at large $Q^2$ with $t=(p-p')^2$, $x_{Bj}=Q^2/2(pq)$ and
$q'^2=M_F^2$  fixed, are amenable to QCD description. The
factorization theorem \cite{CFS} asserts that the amplitude
of the process (\ref{proc1}) has the following form:

\begin{eqnarray}
   &&
   \sum _{i,j} \int _{0}^{1}dz  \int dx_1\,
   f_{i/T}^{T'}(x_1 ,x_1 -x_{Bj};t,\mu ) \,
   H_{ij}(Q^2 x_1/x_{Bj},Q^{2},z,\mu )
   \, \Phi^F_{j}(z,\mu )
\nonumber\\
&&
   + \mbox{power-suppressed corrections} ,
\label{factorization}
\end{eqnarray}
where $f_{i/T}^{T'}$ is $T\to T'$ skewed parton distribution
(for review and referencies see
\cite{JiReview}), $\Phi^F_{j}(z,\mu )$ is the distribution
amplitude of the hadronic state $F$ (not necessarily one particle
state), and $H_{ij}$ is a hard part computable in pQCD as series in
$\alpha_s(Q^2)$. Here we shall study general properties of the
distribution amplitudes $\Phi^F_{j}(z,\mu )$ when
the final hadronic state $F$ is a two pion state
($F=\pi\pi$).

The $2\pi$DA's were introduced recently in \cite{Ter} in the context of
QCD description of the process $\gamma^\ast \gamma \to 2 \pi$.
In ref.~\cite{PW98,MVP98} it was suggested that
the usages of $2\pi$DA's to describe a hard electroproduction of
two pions leads to universal picture of resonant and non-resonant
two pion production. Below we listed main properties of $2\pi$DA's
and shortly discuss their applications to hard diffractive production
of two pions off nucleon.

\vspace{0.1cm}
\noindent
\underline{{\bf $\bullet$   Definition}}:

\be \Phi^{ab} (z, \zeta, m_{\pi\pi}^2 )
&=& \frac{1}{4\pi} \int dx^- e^{-\frac{i}{2}z P^+ x^-}
{}_{out}\langle
\pi^a(p_1) \, \pi^b(p_2) \, | \, \bar\psi(x) \; \hat n \, T \; \psi(0)\,
|0\rangle \Bigr|_{x^+=x_\perp=0} ,
\label{definition1}
\ee
Here, $n$ is a light--like vector ($n^2 = 0$), which we take as
$n_\mu=(1,0,0,1)$. For any vector, $V$, the ``plus'' light--cone
coordinate is defined as $V^+ \equiv (n\cdot V) = V^0 + V^3$; the
``minus'' component as $V^- = V^0 - V^3$. The outgoing pions have
momenta $p_1, p_2$, and $P \equiv p_1 + p_2$ is the total momentum of
the final state. Finally, $T$ is a flavor matrix ($T = 1$ for the
isosinglet, $T = \tau^3/2$ for the isovector $2\pi$DA).
The generalized distribution amplitudes,
eq.~(\ref{definition1}), depend
on the following kinematical variables: the quark momentum fraction with
respect to the total momentum of the two--pion state, $z$; the variable
$\zeta \equiv p_1^+ / P^+$ characterizing the distribution of
longitudinal momentum between the two pions, and the invariant mass,
$m_{\pi\pi}^2 = P^2$.

\vspace{0.1cm}
\noindent
\underline{{\bf $\bullet$
The isospin decomposition}}:

\be
\Phi^{ab}=\delta^{ab}\tr(T)\Phi^{I=0} +\frac12 \tr([\tau^a,\tau^b] T)
\Phi^{I=1}.
\ee

\vspace{0.1cm}
\noindent
\underline{{\bf $\bullet$
Symmetries and normalization}}:
From the C-parity one can easily derive the following symmetry
properties of the
isoscalar ($I=0$) and isovector ($I=1$) parts of $2\pi$DA's
eq.~(\ref{definition1}):

\be
\nonumber
\Phi^{I=0}(z,\zeta,m_{\pi\pi}^2)&=&
-\Phi^{I=0}(1-z,\zeta,m_{\pi\pi}^2)=\Phi^{I=0}(z,1-\zeta,m_{\pi\pi}^2),\\
\Phi^{I=1}(z,\zeta,m_{\pi\pi}^2)&=&\Phi^{I=1}(1-z,\zeta,m_{\pi\pi}^2)=
-\Phi^{I=1}(z,1-\zeta,m_{\pi\pi}^2)\; .
\label{reflection}
\ee
The isospin one parts of $2\pi$DA's
eq.~(\ref{definition1}) is normalized as follows:
\be
\int_0^1 dz \Phi^{I=1}(z, \zeta, m_{\pi\pi}^2 )
=(2\zeta-1) F_\pi^{{\rm e.m.}}(m_{\pi\pi}^2)\; ,
\label{norm}
\ee
where $F_\pi^{{\rm e.m.}}(m_{\pi\pi}^2)$ is the pion e.m. form factor in the time-like region
($F_{\pi}^{{\rm e.m.}}(0)=1$).

\noindent
For the isoscalar $2\pi$DA  $\Phi^{I=0}$ we have the following
normalization condition:
\be
\nonumber
\int_0^1 dz (2 z-1)\Phi^{I=0}(z, \zeta, m_{\pi\pi}^2 )
=-2\ M_2^{\pi}\zeta(1-\zeta) F_\pi^{{\rm EMT}}(m_{\pi\pi}^2)\; ,
\label{norm1}
\ee
where $M_2^\pi$ is a momentum fraction carried by quarks in
a pion, $F_\pi^{{\rm EMT}}(m_{\pi\pi}^2)$ is a pion form factor of
quark part of
energy momentum tensor normalized by $F_\pi^{{\rm EMT}}(0)=1$.
In ref.~\cite{MVP98} this form factor was estimated in the instanton
model of QCD vacuum at low two-pion invariant mass, the result:

\be
\nonumber
F_\pi^{{\rm EMT}}(m_{\pi\pi}^2)=1+\frac{N_c m_{\pi\pi}^2}{48 \pi^2
f_\pi^2}+\ldots \; ,
 \ee
where $f_\pi =93$~MeV is a pion decay constant.

\vspace{0.1cm}
\noindent
\underline{{\bf $\bullet$
Double decomposition in conformal and partial waves
}}:
Let us decompose $2\pi$DA's in eigenfunctions of the ERBL \cite{ERBL}
evolution equation (Gegenbauer polynomials $C_n^{3/2}(2 z-1)$) and
in partial waves of produced pions
(Gegenbauer polynomials $C_l^{1/2}(2 \zeta-1)$).
Generically the decomposition
(for both isoscalar
and isovector DA's) has the form:

\be
\Phi (z, \zeta, m_{\pi\pi}^2 )=
6z(1-z)\sum_{n=0}^{\infty}
\sum_{l=0}^{n+1} B_{nl}(m_{\pi\pi}^2) C_n^{3/2}(2 z-1)C_l^{1/2}(2
\zeta-1).
\label{razhl} \ee
Using symmetry properties
Eq.~(\ref{reflection}) we see that the index $n$ goes over even (odd) and
index $l$ goes
over odd (even) numbers for isovector (isoscalar) $2\pi$DA's.
Normalization conditions Eq.~(\ref{norm}) correspond to
$B_{01}^{I=1}(m_{\pi\pi}^2)=F_\pi^{{\rm e.m.}}(m_{\pi\pi}^2)$.

\vspace{0.1cm}
\noindent
\underline{{\bf $\bullet$
Relations of $2\pi$DA's to quark distribution in a pion
}}:
By crossing symmetry the $2\pi$DA's are related to so-called
skewed parton distributions (see review \cite{JiReview}).
The latter are defined as:
\be
&&\int \frac{d\lambda }{2\pi }
e^{i \lambda \tau}
\langle \pi^a(p^{\prime })|T \bigl\{
\bar \psi_{f'}
(-\lambda n/2){\hat n}
 \psi_f (\lambda n/2)\bigr\}
|\pi^b(p)\rangle = \\
\nonumber
&&\delta^{ab}\delta_{ff'} H^{I=0}(\tau,\xi,t) +
i\varepsilon^{abc}
\tau^c_{ff'} H^{I=1}(\tau,\xi,t)
\; ,
\ee
where $\xi$ is a skewedness parameter defined as:
$\xi=-(n\cdot (p'-p))/(n\cdot(p'+p))$, $t=(p'-p)^2$ and light-cone
vector $n^\mu$ normalized by $(n\cdot(p'+p))=2$.
By crossing symmetry we can easily express the moments of
skewed distributions to coefficients $B_{nl}$ in the expansion
(\ref{razhl}) (detailed discussion see in \cite{PW99}):
\be
\int_{-1}^1 d\tau \tau^{N-1} H^I(\tau,\xi,t)=
\sum_{n=0}^{N-1}
\sum_{l=0}^{n+1} B^I_{nl}(t) \xi^N
C_l^{1/2}\biggl(\frac{1}{\xi}\biggr)\
\int_{-1}^1 d x
\frac{3}{4}[1-x^2]\ x^{N-1} C_n^{3/2}(x) \, .
\ee
If we take the forward limit in this formula, we obtain the relations
between moments of quark distributions in a pion and coefficients
$B_{nl}$:
\be
\nonumber
M_N^{(\pi)}&\equiv& \int_{0}^1 x^{N-1} (q_\pi(x)-\bar q_\pi(x))=
B^{I=1}_{N-1,N}(0) A_N \qquad \mbox{ for \ odd\ } N\, ,\\
M_N^{(\pi)}&\equiv& 2 \int_{0}^1 x^{N-1} (q_\pi(x)+\bar
q_\pi(x))= B^{I=0}_{N-1,N}(0) A_N \qquad \mbox{for \ even\ } N \, ,
\label{f2tp}
\ee
where $A_N$ are numerical coefficients ($e.g.$ $A_1=1, A_2=9/5,
A_3=6/7,A_4=5/3$, etc.) and $q_\pi(x)=u^{\pi^+}(x)$.
For example, from eq.~(\ref{f2tp}) we obtain that
$B^{I=1}_{01}(0)=M_1^{(\pi)}=1$ what corresponds to normalization
condition (\ref{norm}). Also it is easy to see that
$B^{I=0}_{12}(0)=5/9 M_2^{(\pi)}$ corresponds to normalization
(\ref{norm1})\footnote{To see this one has to use additionally soft
pion theorem (\ref{letwf0}) for isoscalar $2\pi$DA}.

\vspace{0.1cm}
\noindent
\underline{{\bf $\bullet$
Evolution
}}:
The Gegenbauer polynomials $C_n^{3/2}(2 z-1)$ are eigenfunctions of the
ERBL \cite{ERBL} evolution equation and hence the coefficients
$B_{nl}$ are renormalized multiplicatively (for even $n$ and odd $l$):

\be
B_{nl}(m_{\pi\pi}^2;\mu)=B_{nl}(m_{\pi\pi}^2;\mu_0)
\biggl(
\frac{\alpha_s(\mu)}{\alpha_s(\mu_0)}
\biggr)^{\gamma_n/(2\beta_0)}\;,
\label{renorm}
 \ee
where $\beta_0=11-2/3 N_f$ and the one loop anomalous dimensions are
\cite{VY}:
\be
\nonumber
\gamma_n&=&\frac{8}{3}
\Bigl(1-\frac{2}{(1+n)(2+n)}+4\sum_{k=2}^{n+1}\frac{1}{k} \Bigr) \; .
 \ee

From the decomposition Eq.~(\ref{razhl}) and Eq.~(\ref{renorm}) we can
make a simple prediction for the ratio of the
hard $P-$ and $F-$ waves production amplitudes of
pions in the reaction $\gamma^\ast p\to 2 \pi p$
at asymptotically large virtuality of the incident photon
$Q^2\to\infty$ and fixed $m_{\pi\pi}^2$:

\be
\frac{F{\rm -wave\ amplitude}}{P{\rm -wave\ amplitude}}\sim
\frac{1}{\log(Q^2)^{50/(99-6 N_f)}}\; ,
\ee
or generically for the $2 k+1$ wave:
\be
\frac{(2k+1) {\rm -wave\ amplitude}}{P {\rm -wave\ amplitude}}\sim
\frac{1}{\log(Q^2)^{\gamma_{2k}/(2\beta_0)}}\; .
\ee

\vspace{0.1cm}
\noindent
\underline{{\bf $\bullet$
Soft pion theorems for
two-pion distribution amplitudes}} relate $2\pi$DA's to distribution
amplitude of one pion:

\be
\Phi^{I=1}(z,\zeta=1,m_{\pi\pi}^2=0)=
-\Phi^{I=1}(z,\zeta=0,m_{\pi\pi}^2=0)=\varphi_\pi(z)\, ,
\label{letwf}
\ee
where $\varphi_\pi(z)$ is a one pion DA.
The analogous theorem for an isoscalar part of the $2\pi$DA's has the
form:
 \be \Phi^{I=0}(z,\zeta=1,m_{\pi\pi}^2=0)=
\Phi^{I=0}(z,\zeta=0,m_{\pi\pi}^2=0)=0\; .
\label{letwf0}
\ee
The soft pion theorem Eq.~(\ref{letwf}) allows us to relate
the coefficients $B_{nl}^{I=1}(m_{\pi\pi}^2)$ (see Eq.~(\ref{razhl}))
and the coefficients of expansion of the pion DA in Gegenbauer
polynomials
\be \varphi_\pi(z)=6 z(1-z)\biggl ( 1+\sum_{n={\rm even}}
a_n^{(\pi)} C_n^{3/2}(2 z-1) \biggr).
\label{razhlpi}
\ee
The relation has the form:
\be
a_n^{(\pi)}=\sum_{l=1}^{n+1} B_{nl}^{I=1}(0)\; .
\label{letcoef}
\ee

\vspace{0.1cm}
\noindent
\underline{{\bf $\bullet$
Dispersion relations and their solution for $2\pi$DA's}}:
The $2\pi$ DA's are generically  complex functions
due to the strong interaction of the produced pions.
Above the two-pion threshold $m_{\pi\pi}^2=4 m_\pi^2$
the $2\pi$DA's develop the imaginary
part corresponding to the contribution of on-shell intermediate states
($2\pi$, $4\pi$, etc.). In the region $m_{\pi\pi}^2<16 m_\pi^2$ the imaginary part
is related to the pion-pion scattering amplitude by Watson theorem
\cite{Watson}.
This relation can be written in the following form (see \cite{MVP98}):

\be
\mbox{Im}\, B_{nl}^I(m_{\pi\pi}^2)=
\sin[\delta_l^I(m_{\pi\pi}^2)]e^{i\delta_l^I(m_{\pi\pi}^2)}
B_{nl}^I(m_{\pi\pi}^2)^\ast=
\tan[\delta_l^I(m_{\pi\pi}^2)]\mbox{Re}B_{nl}^I(m_{\pi\pi}^2)\; .
\label{imbnl}
\ee
Using  Eq.~(\ref{imbnl}) one
can write an $N$-subtracted dispersion relation for the $B_{nl}^I(m_{\pi\pi}^2)$

\be
B_{nl}^I(m_{\pi\pi}^2)=\sum_{k=0}^{N-1} \frac{m_{\pi\pi}^{2k}}{k!}\frac{d^k}{dm_{\pi\pi}^{2k}}
B_{nl}^I(0)+
\frac{m_{\pi\pi}^{2N}}{\pi}\int_{4m_\pi^2}^\infty
ds \frac{\tan\delta_l^I(s)\,\mbox{Re}B_{nl}^I(s)}{s^N(s-m_{\pi\pi}^2-i0)}\; .
\label{dr}
\ee
Solution of such type dispersion relation was found long ago
in \cite{omnes}, the solution has the exponential form:

\be
B_{nl}^I(m_{\pi\pi}^2)=B_{nl}^I(0)\exp
\biggl\{
\sum_{k=1}^{N-1} \frac{m_{\pi\pi}^{2k}}{k!}\frac{d^k}{dm_{\pi\pi}^{2k}}\log\, B_{nl}^I(0)
+
\frac{m_{\pi\pi}^{2N}}{\pi}\int_{4m_\pi^2}^\infty
ds \frac{\delta_l^I(s)}{s^N(s-m_{\pi\pi}^2-i0)}
\biggr\}\; .
\label{thesolution}
\ee
A great advantage of the solution Eq.~(\ref{thesolution}) is that
it gives the $2\pi$DA's  in a wide range of energies in terms
of known $\pi\pi$ phase shifts and a few subtraction constants
(usually two is sufficient). The key observation is that these
constants (non-perturbative input) are related to the {\it low-energy}
behaviour  of the $2\pi$DA's
at $m_{\pi\pi}^2\to 0$. In the low energy region they
can be computed using the effective chiral lagrangian.

\vspace{0.1cm}
\noindent
\underline{{\bf $\bullet$
Amplitudes of $\pi\pi$ resonances production}}:
We give here an example how $\rho$ meson distribution amplitude
can be expressed in terms of $2\pi$DA. General formula for DA's
of resonances with any spin can be found in \cite{MVP98}.
In \cite{MVP98} we obtained the following
expression for the coefficients of the expansion of $\rho$ meson DA in
Gegenbauer polynomials

$$
\varphi_\rho(z)=6 z(1-z)\biggl(
 1+\sum_{n={\rm even}}
a_n^{(\rho)} C_n^{3/2}(2 z-1) \biggr),
$$
in terms of only subtraction constants entering Eq.~(\ref{thesolution})

\be
a_n^{(\rho)}=B^{I=1}_{n1}(0)
\exp
\bigl(
\sum_{k=1}^{N-1} c_k^{(n1)} m_\rho^{2k}
\bigr)\; ,
\label{rho2pi}
\ee
where the subtraction constants $c_{k}^{(nl)}$ can be expressed
in terms of $B^{I=1}_{nl}(m_{\pi\pi}^2)$ at low energies
\be
c_{k}^{(nl)}= \frac{1}{k!}\frac{d^k}{dm_{\pi\pi}^{2k}}
[\log\, B_{nl}(m_{\pi\pi}^2)-\log\, B_{l-1\,l}(m_{\pi\pi}^2)]\Biggr|_{m_{\pi\pi}^2=0}\;.
\label{ck}
\ee
The normalization constants $f_\rho$ can be computed as:

\be
f_\rho=\frac{\sqrt 2 \Gamma_\rho
\mbox{Im}F_\pi^{{\rm e.m.}}(m_\rho^2)
 }{g_{\rho\pi\pi}}\;.
\label{dominance2}
\ee

Here we quote the
result for the chirally even $\rho$ meson DA extracted from $2\pi$DA
using Eqs.~(\ref{rho2pi},\ref{dominance2}) and values of
low-energy subtraction constants calculated
using effective chiral lagrangian
( see details and results of these calculations in
refs.~\cite{PW98,MVP98}) :

\be
\varphi_\rho(z)=6 z(1-z)\bigl[1-0.14\, C_2^{3/2}(2 z-1)-
0.01\, C_4^{3/2}(2 z-1)+\ldots \bigr]\;,
\label{ourrhowf}
\ee
with normalization constant $f_\rho=190$~MeV according to
Eq.~(\ref{dominance2}) in a good agreement with
experimental value $f_\rho=195\pm 7$~MeV \cite{PDG}.
The $\rho$ meson DA's were the subject
of the QCD sum rules calculations \cite{BB,BM}, our result
Eq.~(\ref{ourrhowf}) is in a qualitative disagreement with the
results of QCD sum rule
calculations, the sign of $a_2$ obtained here
is opposite to the QCD sum rules
predictions $a_2^{(\rho)}=0.18\pm 0.1$ \cite{BB} and
$a_2^{(\rho)}=0.08
\pm 0.02$, $a_4^{(\rho)}=-0.08\pm 0.03 $ \cite{BM} (these results refer
to normalization point $\mu=1$~GeV).
Let us note that actually our sign of $a_2^{(\rho)}$ follows from:
1) soft pion theorem (\ref{letwf}) 2) the relation of $B_{N-1,N}$
to moments of quark distribution in a pion (\ref{f2tp}) 3) relation
$a_2^{(\rho)}=B^{I=1}_{21}(0) \exp(c_1^{(21)}m_\rho^2) $
(see eq.(\ref{rho2pi})) and 4) the fact that $a_2^{(\pi)}\approx 0$.
Combining all this we can state that:
\be
\mbox{sign}(a_2^{(\rho)})=\mbox{sign}(B^{I=1}_{21}(0))=
\mbox{sign}(a_2^{(\pi)}-B^{I=1}_{23}(0))=
\mbox{sign}(a_2^{(\pi)}-\frac 76 M_3^{(\pi)})=
-\mbox{sign}(M_3^{(\pi)})\, .
\ee
Here in the last equality we assumed that $a_2^{(\pi)}<\frac 76
M_3^{(\pi)}$ what is satisfied in the model and seems
phenomenologically.

\vspace{0.1cm}
\noindent
\underline{{\bf $\bullet$
QCD parametrization of two pion spectrum in diffractive
two pion production}}:
The dependence of the two-pion hard production amplitude
 on the $m_{\pi\pi}$ factorizes into the factor:
\be
A\propto \int_0^1\frac{dz}{z(1-z)} \,\Phi^{I=1}(z,\zeta,m_{\pi\pi}^2; \bar
Q^2)\;, \ee
for the two pions in the isovector state, and

\be
A\propto \int_0^1 dz\frac{(2 z-1)}{z(1-z)}\,
\Phi^{I=0}(z,\zeta,m_{\pi\pi}^2; \bar Q^2)\;,
\ee
for the pions in the isoscalar state ($e.g.$ for the
$\pi^0\pi^0$ production).  In the above equations we showed also
the dependence of the $2\pi$DA's on the scale of the process $\bar Q^2$,
which is governed by the ERBL evolution equation \cite{ERBL}.

For the hard exclusive reactions off nucleon at small $x_{Bj}$
(see $e.g.$ recent measurements \cite{zeus}) the production of two
pions in the isoscalar channel is strongly suppressed
relative to the isovector channel because the former is
mediated by $C$-parity odd exchange. At asymptotically large $Q^2$
one can use the asymptotic form of isovector $2\pi$DA:

\be
\lim_{\bar Q^2\to\infty}\Phi^{I=1}(z,\zeta,m_{\pi\pi}^2; \bar Q^2)=
6 F_\pi^{{\rm e.m.}}(m_{\pi\pi}^2) z(1-z) (2\zeta-1)\; ,
\ee
where $F_\pi^{{\rm e.m.}}(m_{\pi\pi}^2)$ is the pion e.m. form factor in the time-like region.
Therefore the shape of
$\pi^+\pi^-$ mass spectrum in the hard electroproduction process
at small $x_{Bj}$ and asymptotically large $Q^2$ should
be determined completely by the
pion e.m. form factor in time-like region:

\be
\lim_{Q^2\to\infty} A \propto e^{i\delta_1^1(m_{\pi\pi}^2)}|
F_\pi^{{\rm e.m.}}(m_{\pi\pi}^2)| (2
\zeta-1) \;.
\label{asymf}
\ee
Deviation of the $\pi^+\pi^-$ mass spectrum from its asymptotic form
Eq.~(\ref{asymf})
(``skewing") can be parametrized at small $x_{Bj}$
and large $\bar Q^2$ in the form:

\be
\nonumber
A\propto e^{i\delta_1^1(m_{\pi\pi}^2)}|
F_\pi^{{\rm e.m.}}(m_{\pi\pi}^2)|\Bigl[
1+B_{21}(0;\mu_0)\exp\{c_1^{(21)} m_{\pi\pi}^2\}
\biggl(
\frac{\alpha_s(\bar Q^2)}{\alpha_s(\mu_0)}
\biggr)^{50/(99-6 N_f)}
\Bigr] (2\zeta-1)+\\
\nonumber
e^{i\delta_3^1(m_{\pi\pi}^2)}B_{23}(0;\mu_0)\exp\{b_{23} m_{\pi\pi}^2+R_3^1(m_{\pi\pi}^2)\}
\biggl(
\frac{\alpha_s(\bar Q^2)}{\alpha_s(\mu_0)}
\biggr)^{50/(99-6 N_f)}
C_3^{1/2}(2\zeta-1)\\
+\mbox{\rm higher\ powers\ of\ }1/\log(\bar
Q^2) \;,
\ee
here
\be
R_l^I(m_{\pi\pi}^2)= \mbox{\rm Re}
\frac{m_{\pi\pi}^{4}}{\pi}\int_{4m_\pi^2}^\infty
ds \frac{\delta_l^I(s)}{s^2(s-m_{\pi\pi}^2-i0)}\ \ {\rm and\ } \
b_{nl}=
\frac{d}{dm_{\pi\pi}^2}
\log\, B_{nl}(m_{\pi\pi}^2)\Biggr|_{m_{\pi\pi}^2=0}\;.
\label{paramf}
\ee
We see that the deviation of the $\pi\pi$ invariant mass spectrum from
its asymptotic form Eq.~(\ref{asymf}) in this approximation is
characterized by a few low-energy constants ($B_{21}(0)$, $B_{23}(0)$,
$c_1^{(21)}$ , $b_{23}$), other quantities- the pion e.m. form factor
and the $\pi\pi$ phase shifts--are known from low-energy experiments.
In principle, using the parametrization (\ref{paramf}) one can extract
the values of these low-energy parameters from the shape of $\pi\pi$
spectrum (not absolute cross section!) in diffractive production
experiments. Knowing them one can obtain the deviation of the $\pi$
meson DA (see Eq.~(\ref{letcoef}))
\be \nonumber a_2^{(\pi)}=B_{21}(0)
+ B_{23}(0) \;, \ee
and the $\rho$ meson DA (see Eq.~(\ref{rho2pi}))
\be
\nonumber
a_2^{(\rho)}=B_{21}(0) \exp(c_1^{(21)}m_\rho^2)  \;,
\ee
from their asymptotic form $6z(1-z)$, as well as the normalization
constants for the DA of isovector resonances of spin three.

In analysis of experiments on two pion diffractive production
off nucleon (see e.g. \cite{zeus}) the non-resonant background
is described by S\"oding parametrization \cite{soding}, which
takes into account rescattering of produced pions on final
nucleon. Let us note however that in a case of hard ($Q^2\to \infty$)
diffractive production the final state interaction of pions with
residual nucleon is suppressed by powers of $1/Q^2$ relative to the
leading twist amplitude.
Here we proposed alternative leading-twist parametrization
(\ref{paramf}) describing the so-called ``skewing" of two pion
spectrum.

\vspace{0.3cm}
\noindent
\underline{{\bf
Acknowledgments}}\\
I am grateful to D.~Diakonov, M.~Diehl, L.~Frankfurt, K.~Goeke,
P.~Pobylitsa, A.~Radyushkin,  A.~Sch\"afer, M.~Strikman, O.~Teryaev and
C.~Weiss for inspiring discussions.
This work has been
supported in part by the BMFB (Bonn), Deutsche Forschungsgemeinschaft
(Bonn) and by COSY (J\"ulich).

%
%

%


\begin{thebibliography}{99}
%
\bibitem{MVP98} M.V.~Polyakov, hep-ph/9809483.
\bibitem{CFS}  J.~Collins,L.~Frankfurt, and M.~Strikman,
 Phys. Rev. {\bf D56} (1997) 2982.
\bibitem{Rad} A.~V. Radyushkin, Phys. Rev. {\bf D56} (1997)
5524.
\bibitem{JiReview}
X. Ji, J. Phys. {\bf G24} (1998) 1181.

\bibitem{Ter} M.\ Diehl, T.\ Gousset, B.\ Pire and O.\ Teryaev,
Phys. Rev. Lett. {\bf 81} (1998) 1782.

\bibitem{PW98} M.V.\ Polyakov and C. Weiss, Bochum University
preprint RUB-TPII-7/98, hep-ph/9806390, Phys. Rev. D in press.
\bibitem{zeus} ZEUS Collaboration, hep-ex/9808020.


\bibitem{ERBL}
G.P.~Lepage and S.J.~Brodsky, Phys.\ Lett.\ {\bf B 87} (1979) 359;
A.V.~Efremov and A.V.~Radyushkin, Phys.\ Lett.\ {\bf B 94} (1980) 245.

\bibitem{VY}
M.A.\ Shifman and M.I.\ Vysotsky,
Nucl.\ Phys.\ {\bf B 186}, 475 (1981).



\bibitem{Watson} K.M.~Watson, Phys. Rev. {\bf 95} (1955) 228.


\bibitem{omnes}
R.~Omnes, Nuovo Cim. {\bf 8} (1958) 316.

\bibitem{BB} P.~Ball and V.~M.~Braun, Phys. Rev. {\bf D54} (1996)
2186.
\bibitem{BM} A.P.\ Bakulev and S.V.\ Mikhailov,
Preprint JINR-E2-97-419 (1998), hep-ph/9803298.

\bibitem{PDG} Particle Data Group, Phys. Rev. {\bf D54 } (1996), 1.

\bibitem{PW99} M.V.~Polyakov and C.~Weiss,
Bochum University preprint, RUB-TP2-01/99.
\bibitem{soding} P.~S\"oding, Phys. Lett. {\bf B19} (1966) 702.

\end{thebibliography}
\end{document}